
\magnification=1200
\voffset=0 true mm
\hoffset=0 true in
\hsize=6.5 true in
\vsize=8.5 true in
\normalbaselineskip=13pt
\def\doublespace{\baselineskip=20pt plus 3pt\message{double space}}
\def\singlespace{\baselineskip=13pt\message{single space}}
\let\spacing=\singlespace
\parindent=1.0 true cm



\newcount\equationumber \newcount\sectionumber 
\sectionumber=1 \equationumber=1               
\def\setsection{\global\advance\sectionumber by1 \equationumber=1}

\def\numbe{{{\number\sectionumber}{.}\number\equationumber}
                            \global\advance\equationumber by1}

\def\numberit{\eqno{(\number\equationumber)} \global\advance\equationumber by1}

\def\numberal{(\number\equationumber)\global\advance\equationumber by1}

\def\sectionit{\eqno{(\numbe)}}

\def\ccf#1{\,\vcenter{\normalbaselines
    \ialign{\hfil$##$\hfil&&$\>\hfil ##$\hfil\crcr
      \mathstrut\crcr\noalign{\kern-\baselineskip}
      #1\crcr\mathstrut\crcr\noalign{\kern-\baselineskip}}}\,}
\def\scf#1{\,\vcenter{\baselineskip=9pt
    \ialign{\hfil$##$\hfil&&$\>\hfil ##$\hfil\crcr
      \vphantom(\crcr\noalign{\kern-\baselineskip}
      #1\crcr\mathstrut\crcr\noalign{\kern-\baselineskip}}}\,}

\def\small3j#1#2#3#4#5#6{\def\st{\scriptstyle} 
   \bigl(\scf{\st#1&\st#2&\st#3\cr
           \st#4&\st#5&\st#6\cr} \bigr)}




\def\ref#1{$^{#1)}$}    


\def\upa#1{\raise 1pt\hbox{\sevenrm #1}}
\def\dna#1{\lower 1pt\hbox{\sevenrm #1}}
\def\dnb#1{\lower 2pt\hbox{$\scriptstyle #1$}}
\def\dnc#1{\lower 3pt\hbox{$\scriptstyle #1$}}
\def\upb#1{\raise 2pt\hbox{$\scriptstyle #1$}}
\def\upc#1{\raise 3pt\hbox{$\scriptstyle #1$}}
\def\hprime{\raise 2pt\hbox{$\scriptstyle \prime$}}
\def\ccom{\,\raise2pt\hbox{,}}

\def\asymptotically#1{\;\rlap{\lower 4pt\hbox to 2.0 true cm{
    \hfil\sevenrm  #1 \hfil}}
   \hbox{$\relbar\joinrel\relbar\joinrel\relbar\joinrel
     \relbar\joinrel\relbar\joinrel\longrightarrow\;$}}
\def\Asymptotically#1{\;\rlap{\lower 4pt\hbox to 3.0 true cm{
    \hfil\sevenrm  #1 \hfil}}
   \hbox{$\relbar\joinrel\relbar\joinrel\relbar\joinrel\relbar\joinrel
     \relbar\joinrel\relbar\joinrel\relbar\joinrel\relbar\joinrel
     \relbar\joinrel\relbar\joinrel\longrightarrow$\;}}

\catcode`@=11
\def\C@ncel#1#2{\ooalign{$\hfil#1\mkern2mu/\hfil$\crcr$#1#2$}}
\def\gf#1{\mathrel{\mathpalette\c@ncel#1}}      
\def\Gf#1{\mathrel{\mathpalette\C@ncel#1}}      

\def\gapx{\lower 2pt \hbox{$\buildrel>\over{\scriptstyle{\sim}}$}}
\def\lapx{\lower 2pt \hbox{$\buildrel<\over{\scriptstyle{\sim}}$}}

\def\nablaleft{\hbox{\raise 6pt\rlap{{\kern-1pt$\leftarrow$}}{$\nabla$}}}
\def\nablaright{\hbox{\raise 6pt\rlap{{\kern-1pt$\rightarrow$}}{$\nabla$}}}
\def\nablaboth{\hbox{\raise 6pt\rlap{{\kern-1pt$\leftrightarrow$}}{$\nabla$}}}

\def\boks#1#2{{\hsize=#1 true cm\parindent=0pt
  {\vbox{\hrule height1pt \hbox
    {\vrule width1pt \kern3pt\raise 3pt\vbox{\kern3pt{#2}}\kern3pt
    \vrule width1pt}\hrule height1pt}}}}

\def\heading{ }
\def\range{ }

\def\body{\vfill\eject\parindent=1.0 true cm
 \ifx\spacing\singlespace\singlespace\else\doublespace\fi}
\def\title#1{\centerline{{\bf #1}}}

\def\today{\ifcase\month\or
  January\or February\or March\or April\or May\or June\or
  July\or August\or September\or October\or November\or December\fi
  \space\number\day, \number\year}
\let\date=\today
\newcount\hour \newcount\minute
\countdef\hour=56
\countdef\minute=57
\hour=\time
  \divide\hour by 60
  \minute=\time
  \count58=\hour
  \multiply\count58 by 60
  \advance\minute by -\count58

\def\sectionskip{\penalty-500\vskip24pt plus12pt minus6pt}

\def\sec{\hbox{\lower 1pt\rlap{{\sixrm S}}{\hbox{\raise 1pt\hbox{\sixrm S}}}}}
\def\section#1\par{\goodbreak\message{#1}
    \sectionskip\nobreak\noindent{\bf #1}\vskip0.3cm \noindent}

\nopagenumbers
\headline={\ifnum\pageno=\count31\frontheadline
  \else{\ifnum\pageno=0\frontheadline
     \else{{\raise 2pt\hbox to \hsize{\paperhead}}}\fi}\fi}

\footline={\centerline{\sevenbf \folio}}

\def\frontheadline{\sevenbf \hfil}
\def\paperhead{\sevenbf \heading\ \range\hfil\folio}
\newdimen\pagewidth \newdimen\pageheight \newdimen\ruleht
\maxdepth=2.2pt
\pagewidth=\hsize \pageheight=\vsize \ruleht=.5pt

\def\onepageout#1{\shipout\vbox{ 
    \offinterlineskip 
  \makeheadline
    \vbox to \pageheight{
         #1 
 \ifnum\pageno=\count31{\vskip 21pt\line{\the\footline}}\fi
 \ifvoid\footins\else 
 \vskip\skip\footins \kern-3pt
 \hrule height\ruleht width\pagewidth \kern-\ruleht \kern3pt
 \unvbox\footins\fi
 \boxmaxdepth=\maxdepth}
 \advancepageno}}
\output{\onepageout{\pagecontents}}
\count31=-1
\topskip 0.7 true cm
\pageno=0
\doublespace
\centerline{\bf Gauge Theory of Quantum Gravity}
\centerline{\bf J. W. Moffat}
\centerline{\bf Department of Physics}
\centerline{\bf University of Toronto}
\centerline{\bf Toronto, Ontario M5S 1A7}
\centerline{\bf Canada}
\vskip 2 true in
\centerline{\bf December, 1993}
\vskip 2 true in
{\bf UTPT-93-33}
\vskip 0.5 true in
{\bf e-mail: moffat@medb.physics.utoronto.ca}
\par\vfil\eject
\centerline{\bf Gauge Theory of Quantum Gravity}
\centerline{\bf J. W. Moffat}
\centerline{\bf Department of Physics}
\centerline{\bf University of Toronto}
\centerline{\bf Toronto, Ontario M5S 1A7}
\centerline{\bf Canada}
\vskip 1 true in
\centerline{\bf Abstract}
\vskip 0.2 true in
A gauge theory of quantum gravity is formulated, in which
an internal, field dependent metric is introduced which non-linearly
realizes the gauge fields on the non-compact group $SL(2,C)$, while
linearly realizing them on $SU(2)$. Einstein's $SL(2,C)$ invariant
theory of gravity emerges at low energies, since the extra degrees of
freedom associated with the quadratic curvature and the internal metric
only dominate at high energies. In a fixed internal metric gauge, only the
the $SU(2)$ gauge symmetry is satisfied, the particle spectrum
is identified and the Hamiltonian is shown to be bounded from below.
Although Lorentz invariance is broken in this gauge, it
is satisfied in general. The theory is quantized in this fixed,
broken symmetry gauge as an $SU(2)$ gauge theory on a lattice
with a lattice spacing equal to the Planck length. This produces a unitary
and finite theory of quantum gravity.
\par\vfil\eject
\proclaim 1. {\bf Introduction} \par
\vskip 0.2 true in
The problem of quantum gravity continues to be a central issue in modern
physics$^{1}$ and is considered by many to be the greatest challenge in
theoretical physics today. Despite the fact that considerable effort has
been devoted by many physicists over a period of 40 years to solve the
problem of quantum gravity, no significant success has been achieved in
this quest. There have been two views on how to attack the
problem: the first assumes that a correct technical solution
is required, which will unite gravitation theory with
local, relativistically invariant quantum field theory. The second seeks
a new and perhaps radical departure from conventional Einstein gravitational
theory and the axioms of local, relativistic field theory.
In conventional particle theories, it is assumed that the fields propagate
on a non-dynamical, fixed background spacetime, while in modern gravity
theories the spacetime is curved and is the dynamical field , i.e. the
``arena'' plays a central dynamical role in the theory. It could perhaps be
that new methods specially invented to quantize diffeomorphism invariant
theories, like Einstein's theory of gravitation, will succeed in producing
a consistent quantum gravity scheme without the need to change quantum
field theory or Einstein's theory of gravitation. However, it may be also
true that some new idea that changes the conventional picture is needed
to successfully unite gravity and quantum mechanics.

One of the obviously serious drawbacks to discovering the ``right'' quantum
gravity theory is that there is no body of experimental data to guide us in
our search. In the development of early quantum mechanics, there was the
Planck theory of blackbody radiation and the photo-electric effect to
show the way, and the Michelson-Morely experiment was an important
experimental signpost that guided the invention of special relativity.
All we have in our quest for a quantum gravity theory is a perception that
there must exist a ``beautiful'' and mathematically consistent paradigm
that will be accepted by the theoretical physics community as the correct
quantum gravity theory.

Attempts to solve the problem using perturbation theory with
an expansion around a fixed classical background lead either to an
unrenormalizable theory or to a violation of unitarity, or both. The problem
with unitarity is possibly more severe than the lack of renormalizability,
because whereas higher order theories can be found that are renormalizable,
they suffer from ghost poles and lack of unitarity$^{2}$. Moreover, whether
the theory is renormalizable or not is probably irrelevant, for quantum
gravity effects will not become important before the Planck energy
$\sim 10^{19}$ GeV, when the renormalizability and convergence of the
perturbation theory will surely break down above the Planck energy.
A standard field theoretic treatment, based on perturbation theory
using Feynman diagrams obtained from path integrals, or from a canonical
formulation, fails for Einstein's theory at two-loop level and for all
loops when matter is included, and the feature of diffeomorphism
invariance seems to make a non-perturbative approach necessary.
Therefore, it seems imperative that we use non-perturbative methods to
quantize gravity.

Another problem is that standard classical Einstein gravity, based on a
metric and a connection, does not have the form of a classical Yang-Mills gauge
theory. In Einstein's theory of gravity, the metric is the dynamical field,
and the connection is restricted to being a function of the metric by metricity
and torsion-free constraints, while in Yang-Mills theory the connection is the
dynamical variable and the metric is a constant, $\delta_{ab}$. This makes
Einstein's theory appear to be disturbingly different from the Yang-Mills
structure of all other modern field theories, in particular, the standard
model of elementary particles, which has been remarkably successful in its
agreement with
experimental data. Thus, it would be desirable to seek a quantum gravity
theory that is not only consistent, but is also easy to unify with the
successful standard model.

A promising quantization scheme was developed by Ashtekar$^{3}$, in
which Einstein's theory of gravity is reformulated in terms of new
variables. This program had some appealing features that made it
suitable for a canonical quantization of the gravitational field using
Dirac's methods of quantization. Emphasis was shifted from the metric
to the connection, which made it possible to obtain polynomial-type
constraints which could be solved. This appeared to remove the serious
technical problems related to the constraints in earlier attempts to quantize
canonical formulations of gravity. Recently, Capovilla, Dell and Jacobson$^{4}$
have gone even further and developed a theory almost purely in terms of
the connection. However, other serious problems
arose, for inner products of states in the Hilbert space of physical
solutions could not be found and observables could not be obtained
to calculate physical quantities. Ashtekar's Hamiltonian is complex and
reality conditions have to be imposed to extract a real Einstein gravity
theory. In effect, this means that not all the Hamiltonian constraints
can be solved. Without a solution to this problem and the metric-signature
condition, one cannot calculate any physical quantities in the theory.

Attempts have also been made to
develop gauge theories of gravity, path integral quantization formulations
in Euclidean space, and numerical calculations in simplicial lattice space
gravity. These routes to a theory suffered problems with the inherent
non-compact nature of the spacetime group underlying conventional Einstein
gravity and its various gauge theory generalizations.

In the following, we shall develop a new quantum gravity theory by
introducing a different physical principle for small-distance gravity.
Thus, we adopt the position that some new physical insight is needed to
understand the short-distance behavior of gravity, i.e. ``new physics''
at short distances is required to solve quantum gravity.
We shall spontaneously violate local Lorentz invariance in a fixed,
internal metric gauge, while
maintaining a causal theory. In general, the theory is invariant under Lorentz
and diffeomorphism transformations.  By taking this apparently radical step,
we solve many
of the current vexing problems in quantum gravity, without violating
well-known experimental tests of classical general relativity.
Since we only violate Lorentz invariance in the fixed internal metric gauge,
and because there is nothing physically special about this gauge, we can
quantize the gravitational theory as an $SU(2)$ gauge theory
on a lattice with the lattice spacing $a=G^{1/2}$. In this way we obtain a
finite and unitary solution for quantum gravity.

Einstein's classical theory
of gravity emerges in a low-energy limit, since a coupling constant and a
mass control the quadratic gauge piece and the internal
metric contributions to the Lagrangian density, and set a high-energy scale
of the order of the Planck energy. Any corrections to Einstein's
gravitational
theory are down by powers of the inverse Planck mass $\sim E/M_P$. Thus,
Einstein's theory is treated as an ``effective'' theory, valid at macroscopic
distances and is largely independent of small-distance behavior. This point
of view is in accord with the modern treatment of quantum field theory,
which accepts non-renormalizable interactions suppressed by inverse powers
of the cutoff$^{5}$.

We shall quantize the $SU(2)$ version of the Yang-Mills theory
by using non-\break perturbative loop representation methods$^{6}$, which can
be
converted to a lattice gauge theory with a length scale $a$ equal to the
Planck length. Arguments have been put forward$^{7}$ which suggest that the
loop representation applied to quantum gravity naturally leads to
$a$ being equal to the Planck length. This will pave the way for a finite
quantum gravity theory, which in our case will be unitary, as well.

A Higgs-type spontaneous symmetry breaking can also be invoked at Planck
temperatures through a first-order phase transition. Spontaneous violation
of local Lorentz and diffeomorphism
invariance has interesting consequences for the problem of time
in quantum cosmology$^{8}$, early Universe cosmology$^{9}$,
as well as for the problem of information loss in black hole evaporation
$^{10}$.
\vskip 0.2 true in
\setsection\proclaim 2. {\bf SL(2,C) Spinor Gauge Formalism} \par
\vskip 0.2 true in
We shall begin by reviewing the basic properties of $SL(2,C)$
gauge theory of gravity based on a spinor formalism$^{11-13}$.
We define at each point of the four-dimensional spacetime manifold, a complex
two-dimensional linear space, called the spinor space. The elements of
the spinor space are composed of two-dimensional spinors, namely,
two-component complex quantities $\psi^A$, A=0,1.

Let $p^A_a$ be a normalized spinor basis, i.e. any pair of spinors (dyad):
$p^A_a=(p^A_0,p^A_1)$, such that
$$
p^A_a\epsilon_{AB}p^B_b=\epsilon_{ab},\quad
p^A_a\epsilon^{ab}p^B_b=\epsilon^{AB},
\sectionit
$$
where the dyad indices $a,b=0,1$. Also,
$$
\epsilon_{AB}=\epsilon^{AB}=\epsilon_{A'B'}=\epsilon^{A'B'}
=\left(\matrix{0&1\cr
               -1&0\cr}\right),
\sectionit
$$
where a prime on a suffix denotes the complex conjugate operation. The
$\epsilon_{AB}$ satisfy
$$
\epsilon_{AC}\epsilon^{BC}={\delta_A}^B.
\sectionit
$$
Spinor indices are raised and lowered according to the rules:
$$
\psi_A=\psi^B\epsilon_{BA},\quad \psi^A=\epsilon^{AB}\psi_B.
\sectionit
$$
A spinor $\psi^A$ can be expanded in terms of the dyad $p^A_a$, namely
$$
\psi^A=\psi^a p^A_a,
\sectionit
$$
where the scalars $\psi^a$ denote the dyad components of $\psi^A$.

The spinor $\psi_A$ satisfies the transformation law:
$$
\psi_A=\Lambda^B_A\psi^\prime_B,\quad \psi^{\prime A}=\psi^B\Lambda_B^A,
\sectionit
$$
such that
$$
\psi_A\psi^A=\psi^\prime_A\psi^{\prime A},
\sectionit
$$
is an invariant. Here, $\Lambda^B_A$ denotes the spinor transformation
matrix in the transformation law (an element of the group $SL(2,C)$).
The spinors are scalars with respect to the group of general coordinate
transformations in the base manifold.

The covariant differentiation operator for spinors, which allows spinors to
be compared at different spacetime points, is defined by
($\psi_{,\mu}=\partial/\partial x^\mu$):
$$
D_\mu\psi_A=\psi_{A,\mu}-\Omega^B_{A\mu}\psi_B,\quad
D_\mu\psi^A=\psi^A_{,\mu}+\Omega^A_{B\mu}\psi^B,
\sectionit
$$
where $\Omega^B_{A\mu}$ ($\mu=0,1,2,3$) are four two-by-two complex matrices,
denoting the spinor connections, which are subject to the transformation law:
$$
\Omega^{\prime B}_{A\mu}=(\Lambda^{-1})^D_A\Omega^C_{D\mu}\Lambda^B_C
+(\Lambda^{-1})^C_{A,\mu}\Lambda^B_C.
\sectionit
$$
The spinor connections are traceless matrices: $\Omega^A_{A\mu}=0$.
The metric structure of spacetime and the spinor space can be connected by
introducing a Hermitian spinor vector, a set of four Hermitian two-by-two
matrices:
$$
\sigma^\mu_{AB'}=\bar \sigma^\mu_{B'A},
\sectionit
$$
which satisfy the orthogonality conditions:
$$
\sigma^\mu_{AB'}\sigma^{AB'}_\nu={\delta^\mu}_\nu,\quad
\sigma^\mu_{AB'}\sigma^{CD'}_\mu={\delta_A}^C{\delta_{B'}}^{D'}.
\sectionit
$$
Using these algebraic conditions, we obtain the relation:
$$
\sigma^\mu_{AC'}\sigma^{\nu BC'}+\sigma^\nu_{AC'}\sigma^{\mu BC'}
=g^{\mu\nu}{\delta_A}^B,
\sectionit
$$
and
$$
g_{\mu\nu}=\sigma_{\mu AC'}\sigma_\nu^{AC'},
\sectionit
$$
where $g_{\mu\nu}$ is the spacetime pseudo-Riemannian metric tensor. In
flat spacetime with $g_{\mu\nu}=\eta_{\mu\nu}$, where $\eta_{\mu\nu}=
\hbox{diag}(1,-1,-1,-1)$, we can choose the $\sigma$'s as the three
Pauli spin matrices and the unit matrix:
$$
\sigma^0_{AB'}={1\over \sqrt{2}}\left(\matrix{1&0\cr0&1\cr}\right),
\quad \sigma^1_{AB'}={1\over \sqrt{2}}\left(\matrix{0&1\cr1&0\cr}
\right),
$$
$$
\sigma^2_{AB'}={1\over \sqrt{2}}\left(\matrix{0&i\cr-i&0\cr}\right),
\quad \sigma^3_{AB'}={1\over \sqrt{2}}\left(\matrix{1&0\cr0&-1\cr}\right).
\sectionit
$$

We impose the conditions:
$$
D_\nu\sigma^\mu_{AB'}=0,
\sectionit
$$
which determine the relation between the spinor connection $\Omega_\mu$ and
the Christoffel connection $\Gamma^\lambda_{\mu\nu}$:
$$
\Omega^C_{A\mu}={1\over 2}\sigma^{CB'}_\nu (\sigma^\rho_{AB'}
\Gamma^\nu_{\rho\mu}+\sigma^\nu_{AB',\mu}).
\sectionit
$$

We introduce the representation:
$$
D_\mu p^A_a=A^b_{a\mu}p^A_b.
\sectionit
$$
The coefficients $A_{a\mu}^b$, which form a set of four $2\times 2$
complex matrices, will be taken to be the connections of the gauge
theory. They satisfy the traceless condition: $A^a_{a\mu}
=0$ and the transformation law:
$$
A^{\prime d}_{c\mu}=(\Lambda^{-1})^a_c A^b_{a\mu}\Lambda^d_b-
(\Lambda^{-1})^a_c\Lambda^d_{a,\mu}.
\sectionit
$$
For a normalized spinor basis: $p^A_0=(1,0),\,
p^A_1=(0,1)$, we have
$$
A^b_{a\mu}=\Omega^C_{A\mu}p^A_a p^b_C.
\sectionit
$$

The gauge field (curvature) is given by
$$
G^b_{a\mu\nu}=A^b_{a\mu,\nu}-A^b_{a\nu,\mu}+A^c_{a\mu}A^b_{c\nu}
-A^c_{a\nu}A^b_{c\mu}.
\sectionit
$$
We shall adopt the matrix notation:
$$
G_{\mu\nu}=A_{\mu,\nu}-A_{\nu,\mu}+[A_\mu,A_\nu].
\sectionit
$$
The gauge field has the usual transformation law under spinor transformations:
$$
G^{\prime b}_{a\mu\nu}=(\Lambda^{-1})^c_aG^d_{c\mu\nu}\Lambda^b_d.
\sectionit
$$
We also have that
$$
D_\nu D_\mu p^A_a-D_\mu D_\nu p^A_a=G^b_{a\mu\nu}p^A_b
\sectionit
$$
and the traceless condition: $G^A_{A\mu\nu}=0$.
For an arbitrary spinor $\psi^A$, we have
$$
D_\nu D_\mu\psi^A-D_\mu D_\nu\psi^A =(\Omega^A_{B\mu,\nu}-
\Omega^A_{B\nu,\mu}
+\Omega^C_{B\mu}\Omega^A_{C\nu}-\Omega^C_{B\nu}\Omega^A_{C\mu})\psi^B.
\sectionit
$$
This yields
$$
G^B_{A\mu\nu}=(\Omega_{\mu,\nu}-
\Omega_{\nu,\mu}+[\Omega_\mu,\Omega_\nu])^B_A.
\sectionit
$$
The Bianchi identities take the form:
$$
\epsilon^{\alpha\beta\gamma\delta}(G_{\beta\gamma,\delta}+
[G_{\beta\gamma},\Omega_\delta])^B_A=0.
\sectionit
$$

The Riemann curvature tensor is related to the gauge field by the equation:
$$
{R^\alpha}_{\beta\gamma\delta}=-(G^C_{A\gamma\delta}\sigma^\alpha_{CB'}
+G^{C'}_{B'\gamma\delta}\sigma^\alpha_{AC'})\sigma^{AB'}_\beta,
\sectionit
$$
and
$$
G^B_{A\gamma\delta}=-{1\over
2}{R^\alpha}_{\beta\gamma\delta}\sigma_\alpha^{BC'}
\sigma^\beta_{AC'},
\sectionit
$$
where
$$
{R^\delta}_{\alpha\beta\gamma}=\Gamma^\delta_{\alpha\gamma,\beta}-
\Gamma^\delta_{\alpha\beta,\gamma}
+\Gamma^\mu_{\alpha\gamma}\Gamma^\delta_{\mu\beta}-
\Gamma^\mu_{\alpha\beta}\Gamma^\delta_{\mu\gamma},
\sectionit
$$
is the Riemann curvature tensor.
\vskip 0.2 true in
\setsection\proclaim 3. {\bf The Lagrangian Density} \par
\vskip 0.2 true in
We must now formulate a Lagrangian density which guarantees that the
Hamiltonian is bounded from below and that there are no ghost states and
violations of unitarity. Since we are trying to construct an $SL(2,C)$
gauge theory of gravity, we are confronted with the usual
problems of negative probabilities and negative energy when using a
non-compact group$^{14,15,8}$. This may be avoided by using an internal metric
tensor $s_{ab}$, which is Hermitian, non-negative and is a matrix of scalar
fields$^{16-20}$. This leads us to the Lagrangian density:
$$
{\cal L}={\cal L}_1+{\cal L}_2+{\cal L}_3+{\cal L}_M,
\sectionit
$$
where
$$
{\cal L}_1=-{1\over \kappa}\sigma Tr(\Sigma^{\mu\nu}
G_{\mu\nu}),
\sectionit
$$
$$
{\cal L}_2=-\sigma\biggl\{{1\over 4\alpha}Tr(s^{-1}G^{{\dag}\mu\nu}sG_{\mu\nu})
+{m^2\over 8}Tr[(s^{-1}\nabla_\mu s)(s^{-1}\nabla^\mu s)]\biggr\},
\sectionit
$$
$$
{\cal L}_3=\sigma\biggl\{\nabla_\mu\phi^{\dag} s\nabla^\mu\phi
-V(\phi^{\dag} s\phi)\biggr\},
\sectionit
$$
where
$$
\sigma\equiv \sqrt{-g}=[-\hbox{det}(\sigma_{\mu ab'}\sigma^{ab'}_\nu)]^{1/2},
\sectionit
$$
with $g=\hbox{det}(g_{\mu\nu})$, and
$$
\Sigma^b_{a\mu\nu}={1\over 2}(\sigma_{\mu ac'}\sigma^{bc'}_\nu
-\sigma_{\nu ac'}\sigma_\mu^{bc'}).
\sectionit
$$
Moreover, $\kappa=8\pi G$, $\alpha$ is a dimensionless coupling
constant, $m$ is a mass and ${\cal L}_M$ is the matter Lagrangian density.
We have also included a coupling to two complex scalar fields $\phi$, where
$\phi$ transforms as
$$
\phi^\prime=\Lambda^{\dag}\phi.
\sectionit
$$
Moreover, $V$ is a potential which is a function of the $SL(2,C)$
invariant quantity $\phi^{\dag}s\phi$.

The covariant derivative of $s$ with respect to the gauge potential
$A_\mu$ is defined by
$$
\nabla_\lambda s=s_{,\lambda}-A_\lambda^{\dag} s-sA_\lambda,
\sectionit
$$
and the internal metric $s$ transforms as
$$
s^\prime=\Lambda^{\dag} s\Lambda,
\sectionit
$$
where $s^{\dag}$ is the Hermitian conjugate of $s$.

The Lagrangian density ${\cal L}$ is invariant under $SL(2,C)$ gauge
transformations and, as we shall see in the next section, the Hamiltonian
with the correct constraints imposed on it is bounded from below.

By using the anti-commutation relations of the $\sigma$'s, we get
$$
Tr(\Sigma^{\mu\nu}\Sigma_{\alpha\beta})
={1\over 2}(\delta^\mu_\beta\delta^\nu_\alpha
-\delta^\mu_\alpha\delta^\nu_\beta)
+{1\over 2}i{\epsilon^{\mu\nu}}_{\alpha\beta}.
\sectionit
$$
Suppressing dyad indices, we can write Eq. (2.30) in the form:
$$
G_{\mu\nu}={1\over 2}\Sigma^{\alpha\beta}R_{\mu\nu\alpha\beta}.
\sectionit
$$
Substituting this expression into (3.2), using (3.10) and the cyclic identity
obeyed by the curvature tensor:
$$
\epsilon^{\mu\nu\alpha\beta}R_{\mu\nu\alpha\beta}=0,
\sectionit
$$
we get the Einstein-Hilbert Lagrangian density:
$$
{\cal L}_1={1\over 2\kappa}\sqrt{-g}R,
\sectionit
$$
where $R_{\alpha\beta}={R^\sigma}_{\alpha\sigma\beta}$ and
$R={R^\alpha}_\alpha$ are the Ricci tensor and the Ricci scalar, respectively.
The Lagrangian density (3.3) is quadratic in the Riemann curvature tensor.
The most general Lagrangian density involving
quadratic curvature invariants will contain 16 curvature invariants-- 4 Weyl
invariants, 4 Ricci invariants and 8 mixed invariants. Here,
$$
C_{\alpha\beta\gamma\delta}=R_{\alpha\beta\gamma\delta}
-{1\over 2}(g_{\alpha\gamma} R_{\beta\delta}+g_{\beta\delta}R_{\alpha\gamma}
-g_{\alpha\delta}R_{\beta\gamma}-g_{\beta\gamma}R_{\alpha\delta})
$$
$$
+ {1\over
6}(g_{\alpha\gamma}g_{\beta\delta}R-g_{\alpha\delta}g_{\beta\gamma}R),
\sectionit
$$
is the Weyl conformal tensor which satisfies:
$$
C^\alpha_{\beta\alpha\delta}=0.
\sectionit
$$
We have chosen the form of ${\cal L}_2$ by analogy with Yang-Mills theory.
\vskip 0.2 true in
\setsection\proclaim 4. {\bf Classical Hamiltonian} \par
\vskip 0.2 true in
The Hamiltonian constraint formalism and canonical formalism for an
$SL(2,C)$ gauge invariant Lagrangian density of the type we propose for
${\cal L}_2$ in (3.3), has
been analyzed in detail by Dell, deLyra and Smolin$^{19}$ and Popovi\'c$^{20}$.

We shall consider the matter-free case, $T_{\mu\nu}=0$, and we shall
also choose $\phi_\mu=0$. The Lagrangian becomes:
$$
{\cal L}_G={\cal L}_1+{\cal L}_2.
\sectionit
$$
The Lagrangian density (4.1) is independent
of $\hbox{det}(s)$ and we can therefore choose $\hbox{det}(s)=1$. Also,
$s$ has no singularities and we are able to choose a gauge transformation
$q$ such that $s=I$ everywhere:
$$
s^\prime=q^{\dag}sq=I,
\sectionit
$$
where $I$ is the unit matrix. Since $s$ is a Hermitian matrix with unit
determinant, the condition (4.2) fixes the $SL(2,C)/SU(2)$ part of the gauge
freedom, and one is left only with the $SU(2)$ gauge invariance.
In the gauge (4.2), the particle content of the theory becomes manifest.
The $SU(2)$ gauge field is coupled to a multiplet of massive spin-one
vector bosons in the adjoint representation, with the couplings chosen so that
the theory can be extended by means of the metric $s$ to be invariant under
the larger group $SL(2,C)$, which is isomorphic to the homogeneous Lorentz
group $SO(3,1)$. Thus, the Hermitian metric field $s$ has
conspired with the $SL(2,C)/SU(2)$ gauge field to produce massive vector
fields,
while the gauge fields associated with the compact generators of $SU(2)$
remain massless. Although local Lorentz invariance is broken in the
gauge (4.2), there is nothing physically special about this gauge and
in general Lorentz invariance is satisfied. Thus, the Lorentz
symmetry breaking is a special representation of the physical theory, and
the vacuum state is not broken in any gauge, including the one determined
by (4.2). We can therefore choose to quantize the theory in this gauge
and not violate any physical principles.

The coupling constant $\alpha$ and the mass $m$ are
chosen so that at low energies the Lagrangian density ${\cal L}_1$ dominates
and yields $SL(2,C)$ and diffeomorphism invariant Einstein gravity, i.e.,
the extra degrees of freedom associated with the quadratic pieces and
the internal metric $s$ dominate at high energies. At
energies of the order of the Planck mass, where quantum gravity
becomes important, the Lagrangian density in the gauge $s=I$ is broken down to
an $SU(2)$ invariant gauge theory, which approximates classical Einstein
gravity at low energies.

The traceless $2\times 2$ complex matrices $X$ which describe the
$SL(2,C)$ generators can be decomposed into anti-Hermitian and Hermitian
pieces with respect to $s$:
$$
X=i\tau+\lambda,
\sectionit
$$
where
$$
i\tau={1\over 2}(X-s^{-1}X^{\dag}s),\quad \lambda={1\over
2}(X+s^{-1}X^{\dag}s).
\sectionit
$$
We have
$$
(i\tau)=-s^{-1}(i\tau)^{\dag}s,\quad \lambda=s^{-1}\lambda^{\dag}s.
\sectionit
$$
The anti-Hermitian piece $i\tau$ generates the $SU(2)$ maximally compact
subgroup of $SL(2,C)$, which preserves $s$, and which is isomorphic to
the three-dimensional rotation group $SO(3)$. Under an infinitesimal
$\lambda$ transformation:
$$
\delta s=s\lambda+\lambda^{\dag}s.
\sectionit
$$
The $SL(2,C)$ gauge potential $A_\mu$ can be decomposed into
anti-Hermitian and Hermitian pieces with respect to $s\,$$^{19}$:
$$
A_\mu=iV_\mu+B_\mu,
\sectionit
$$
where
$$
iV_\mu={1\over 2}(A_\mu-s^{-1}A^{\dag}_\mu s),\quad
B_\mu={1\over 2}(A_\mu+s^{-1}A^{\dag}_\mu s).
\sectionit
$$
This decomposition is not gauge invariant, for the two pieces mix under
gauge transformations.

The $V_\mu$ fields correspond to the familiar $SU(2)$ Yang-Mills gauge
fields, while the $B_\mu$ fields are related to the $SL(2,C)/SU(2)$
part of the gauge group. We have
$$
\nabla_\mu s=s_{,\mu} - 2sB_\mu.
\sectionit
$$
In general, the Lagrangian density (4.1) is invariant under the full
$SL(2,C)$ group of transformations.

By substituting (4.7) into (3.2), we obtain
$$
{\cal L}_1=
-{1\over \kappa}\sqrt{-g}[Tr(L^{\mu\nu}U_{\mu\nu}+S^{\mu\nu}C_{\mu\nu})],
\sectionit
$$
where
$$
U_{\mu\nu}=V_{\mu,\nu}-V_{\nu,\mu}+[V_\mu,B_\nu]+[B_\mu,V_\nu],
\sectionit
$$
$$
C_{\mu\nu}=b_{\mu\nu}-[V_\mu,V_\nu],
\sectionit
$$
and
$$
b_{\mu\nu}=B_{\mu,\nu}-B_{\nu,\mu}+[B_\mu,B_\nu].
\sectionit
$$
Here, we have used the fact that, in view of (3.13), the Lagrangian density
${\cal L}_1$ is real. This leads to the condition:
$$
Tr(L^{\mu\nu}D_{\mu\nu}+S^{\mu\nu}U_{\mu\nu})=0,
\sectionit
$$
where
$$
D_{\mu\nu}=b_{\mu\nu}+V_{\nu,\mu}-V_{\mu,\nu}.
\sectionit
$$

If we set $s=I$, then the Lagrangian density ${\cal L}_2$, which contains the
quadratic Yang-Mills piece, becomes:
$$
{\cal L}_2=-\sqrt{-g}\biggl[{1\over 4\alpha}Tr(K^{\mu\nu}K_{\mu\nu}+W^{\mu\nu}
W_{\mu\nu})+{m^2\over 2}Tr(B^\mu B_\mu)\biggr],
\sectionit
$$
where
$$
K_{\mu\nu}=f_{\mu\nu}+[B_\mu,B_\nu],\quad W_{\mu\nu}=\tilde\nabla_\mu B_\nu
-\tilde\nabla_\nu B_\mu.
\sectionit
$$
Here, we have
$$
f_{\mu\nu}=V_{\nu,\mu}-V_{\mu,\nu}+[V_\mu,V_\nu],\quad
\tilde\nabla_\mu B_\nu=B_{\nu,\mu}+[V_\mu,B_\nu].
\sectionit
$$

The spontaneous symmetry breaking that occurs here, is the kind associated
with a non-linear realization of the fields on the noncompact group
$SL(2,C)$, inducing a linear realization on the maximal compact subgroup
$SU(2)$$^{16-20}$. This differs from the standard type of Higgs breaking
in that the vacuum expectation values:
$<\phi>_0=<A_\mu>_0=\break <\Omega>_0=0$,
while for the {\it internal} metric $<s>_0=s_0$, we have a non-vanishing
vev. There are no scalar particles left over after the longitudinal parts
of the massive vector fields have been accounted for, and there are no
self-interacting pieces for $s$ in the Lagrangian density, since they are
forbidden by the $SL(2,C)$ gauge invariance. In particular, the ground state
is invariant under the full group of $SL(2,C)$ gauge transformations.

The group $SL(2,C)$ is the double covering group of the homogeneous
Lorentz group $SO(3,1)$ and has the Lie algebra:
$$
Y_m=\sigma_m/2,\quad Z_m=i\sigma_m/2,\quad m=1,2,3,
\sectionit
$$
$$
[Y_m,Y_n]=i\epsilon_{mnk}Y^k,\quad [Y_m,Z_n]=i\epsilon_{mnk}Z^k,\quad
[Z_m,Z_n]=-i\epsilon_{mnk}Y^k,
\sectionit
$$
where the $\sigma_m$ are the three Pauli spin matrices.
Let us use the three-vector notation:
$$
V_m B^m=V\cdot B,\quad \epsilon_{mnk}V^nB^k=(V\times B)_m.
\sectionit
$$
Then, the Lagrangian densities ${\cal L}_1$ and ${\cal L}_2$ can be written:
$$
{\cal L}_1=-{1\over \kappa}\sqrt{-g}(L^{\mu\nu}\cdot U_{\mu\nu}+
S^{\mu\nu}\cdot C_{\mu\nu}),
\sectionit
$$
and
$$
{\cal L}_2=-\sqrt{-g}\biggl[{1\over 4\alpha}(K^{\mu\nu}\cdot
K_{\mu\nu}+W^{\mu\nu}\cdot W_{\mu\nu})+{m^2\over 2}B_\mu\cdot B^\mu\biggr],
\sectionit
$$
where
$$
U_{\mu\nu}=V_{\mu,\nu}-V_{\nu,\mu}+V_\mu\times B_\nu+B_\mu\times V_\nu,
\sectionit
$$
$$
C_{\mu\nu}=b_{\mu\nu}+V_\mu\times V_\nu,
\sectionit
$$
$$
b_{\mu\nu}=B_{\mu,\nu}-B_{\nu,\mu}-B_\mu\times B_\nu.
\sectionit
$$
Moreover, we have
$$
K_{\mu\nu}=V_{\nu,\mu}-V_{\mu,\nu}-V_\mu\times V_\nu+B_\mu\times B_\nu,
\sectionit
$$
$$
W_{\mu\nu}=\tilde\nabla_\mu B_\nu-\tilde\nabla_\nu B_\mu,
\sectionit
$$
$$
\tilde\nabla_\mu B_\nu=B_{\nu,\mu}-V_\mu\times B_\nu.
\sectionit
$$

In the work of Popovi\'c$^{20}$ and Dell, de Lyra and Smolin$^{19}$, a set of
constraint equations was derived using Dirac Hamiltonian constraint
analysis$^{21}$. In a local geodesic (Fermi) frame of coordinates, we have
$g_{\mu\nu}=\eta_{\mu\nu}$, and the canonical Hamiltonian for ${\cal L}_2$
is defined by
$$
H_2=\pi^i_m\dot V^m_i+P^i_m\dot B^m_i - {\cal L}_2,
\sectionit
$$
where
$$
\pi^0_m={\partial {\cal L}_2\over \partial \dot V^m_0}=0,\quad
P^0_m={\partial {\cal L}_2\over \partial \dot B^m_0}=0,
\sectionit
$$
$$
\pi^i_m={\partial {\cal L}_2\over \partial \dot V^m_i}=-{1\over
\alpha}K^{0i}_m,
\sectionit
$$
$$
P^i_m={\partial {\cal L}_2\over \partial \dot B^m_i}=-{1\over \alpha}W^{0i}_m.
\sectionit
$$
The primary constraints are given by (4.31), while the secondary constraints
are of the form:
$$
\{\pi^m_0,H_2\}=C^m=\pi^i_{m,i}+\epsilon^{mnk}\pi^i_n V^k_i+\epsilon^{mnk}P^i_n
B^k_i \approx 0,
\sectionit
$$
$$
\{P^m_0,H_2\}=D^m=P^i_{m,i} - \epsilon^{mnk}P^i_k V^n_i - \epsilon^{mnk}\pi^i_n
B^k_i + m^2 B^m_0 \approx 0.
\sectionit
$$
The $C^m$ generate the usual $SU(2)$ gauge transformations, while the $D^m$
correspond to the $SL(2,C)/SU(2)$ gauge symmetry of the original theory.

When the Dirac brackets are calculated, we find that all second-class
constraints become strong inequalities. In particular, it follows that
$$
\tilde C_m=0,\quad \pi^0_m=0,
\sectionit
$$
are first-class constraints, where
$$
\tilde C_m\equiv C^m - \epsilon^{mnk} B^n_0 P^k_0,
\sectionit
$$
while
$$
D_m=0,\quad  P^0_m=0,
\sectionit
$$
are second-class constraints.

In the presence of the constraints, we obtain the total Hamiltonian:
$$
H=H_1+H_2,
\sectionit
$$
where
$$
H_2={\alpha\over 2}[(\pi^i_m)^2+(P_m^i)^2]+{m^2\over 2}[(B_i^m)^2
+(B_0^i)^2]+{1\over 4\alpha}[(K^m_{ij})^2+(W^m_{ij})^2] \ge 0,
\sectionit
$$
and $H_1$ is the constrained Hamiltonian obtained from the Lagrangian density
(4.22). The total Hamiltonian $H$ is
positive and bounded from below and defines a consistent Hamiltonian system. A
potential
problem with the mass term $B^\mu B_\mu$, due to the indefiniteness of the
Minkowski metric $\eta_{\mu\nu}$, is avoided, because the constraints switch
the sign of the term $m^2B_0^2$ in the Hamiltonian $H_2$. Dell, de Lyra and
Smolin $^{19}$ have shown
in a gauge independent analysis that $H_2$ is indeed ghost free and
tachyonic free and that the boundedness of the Hamiltonian $H_2$ from below is
a
general result in the $SL(2,C)$ (Lorentz) invariant theory.
\vskip 0.2 true in
\proclaim 4. {\bf Non-pertubative Lattice Quantization} \par
\vskip 0.2 true in
The experience of the last thirty or more years of attempts to construct a
consistent quantum gravity theory have shown that a perturbative
expansion around a fixed classical background spacetime cannot succeed
in describing the quantum behavior of the gravitational field. Thus, we have
to resort to a non-perturbative approach to quantize the gravitational
field. The perturbative field theory based on the Lagrangian density (3.1)
does not lead to a renormalizable field theory$^{19}$. Possible additional
dimension four contributions can be added to (3.1), which make the perturbative
theory renormalizable, but at the cost of violating unitarity$^{22}$. At any
rate, even if the theory was perturbatively renormalizable, the perturbation
expansion and the renormalizability will be expected to break down above
the Planck energy, and render the whole scheme useless.
One possible approach to constructing a non-perturbative theory is to use a
loop
representation method$^{6}$
to quantize the gravitational gauge field theory, described by the
Lagrangian density. We shall use this method in conjunction with the
Wilson-type lattice gauge approach$^{23,24}$, in which both space and time
are discretized in the Euclidean version of the theory with a lattice spacing
$a$, which we choose to be equal to the Planck length, $a=G^{1/2}$.

The lattice site will be denoted by a four-vector n. The four-dimensional
integration will be replaced by a sum:
$$
\int d^4x\rightarrow a^4\Sigma_n.
\sectionit
$$
We introduce the non-integrable phase factor of a wavefunction $\psi(x)$:
$$
U(x,x^\prime)=\hbox{exp}\biggl\{i\int_C Y\cdot V_\mu dy^\mu\biggr\},
\sectionit
$$
associated with the gauge potential $V_\mu$. Thus, for the gauge function:
$$
\Phi(\theta)=\hbox{exp}[iY\cdot \theta(x)],
\sectionit
$$
we have
$$
\psi(x)\rightarrow \Phi(\theta)\psi(x),
\sectionit
$$
$$
\bar\psi(x)\rightarrow \bar\psi(x)\Phi^{\dag}(\theta)
$$
and
$$
{U^\prime}(x,x^\prime)=\Phi(\theta)U(x,x^\prime)\Phi^{\dag}(\theta).
\sectionit
$$
The lattice versions of these gauge transformations are:
$$
\psi_n\rightarrow \Phi_n\psi_n,
\sectionit
$$
$$
{\bar\psi}_n\rightarrow {\bar\psi}_n\Phi^{\dag}_n,
\sectionit
$$
and
$$
U(n+\hat \mu,n)^\prime=\Phi_{n+\hat \mu}U(n+\hat \mu,n)\Phi_n^{\dag},
\sectionit
$$
where
$$
\Phi_n=\hbox{exp}(iY\theta_n).
\sectionit
$$
Here, $\hat\mu$ is a four-vector of length $a$ in the direction of $\mu$.

The action for our lattice gauge gravity theory is given by
$$
S_G=\Sigma_p (-g_n)^{1/2}\biggl\{-{1\over \kappa}
(L^{n,\mu\nu}\cdot U_{n,\mu\nu}+S^{n,\mu\nu}\cdot C_{n,\mu\nu})
+{1\over 2\alpha}\{Tr[\hbox{exp}(ia^4 K_{n,\mu\nu})]
$$
$$
+Tr[\hbox{exp}(ia^4 W_{n,\mu\nu})]\}
+{m^2\over 2}Tr[\hbox{exp}(ia^4 B_{n\mu})]\biggr\},
\sectionit
$$
where the sum is over all the plaquettes of the lattice. Moreover, we have
$$
\Sigma_{n,\mu\nu}=iL_{n,\mu\nu}+S_{n,\mu\nu},
\sectionit
$$
where
$$
\Sigma^b_{an,\mu\nu}={1\over 2}(\sigma_{n,\mu ac'}\sigma_{n,\nu}^{bc'}-
\sigma_{n,\nu ac'}\sigma_{n,\mu}^{bc'}).
\sectionit
$$
Also, we have
$$
K_{n,\mu\nu}=\partial_\mu V_{n\nu}-\partial_\nu V_{n\mu}
- V_{n\mu}\times V_{n\nu}+B_{n\mu}\times B_{n\nu},
\sectionit
$$
$$
W_{n,\mu\nu}=\partial_\mu B_{n\nu}-\partial_\nu B_{n\mu}
+V_{n\nu}\times B_{n\mu}-V_{n\mu}\times B_{n\nu},
\sectionit
$$
where
$$
\partial_\mu V_{n\nu}={1\over a}(V_{(n+\hat\mu)\nu}-V_{n\nu})
\sectionit
$$
and $g_n$ and $V_{n\mu}$ denote the spacetime metric tensor and the
gauge potential field at the site $n$. We can recover the
continuum limit from expressions of the form:
$$
{1\over 2}\Sigma_p\biggl\{1-{a^4\over 2}K^{\mu\nu}\cdot K_{\mu\nu}+...\biggr\}
\rightarrow -{1\over 4}\int d^4x K^{\mu\nu}\cdot K_{\mu\nu},
\sectionit
$$
where we have used the tracelessness of the $SU(2)$ matrices and
$Tr(Y^a Y^b)={1\over 2}\delta^{ab}$.

We are now in a position to compute the functional generator:
$$
Z=\int d[V_{n\mu},g_n, B_{n\mu}]{\cal
M}(V_{n\mu},g_n,B_{n\mu})\hbox{exp}(iS_G),
\sectionit
$$
where ${\cal M}$ is a lattice measure. An important question to investigate is
whether there exists a finite lattice formulation of the theory with
$a=G^{1/2}$, since a continuum limit for a non-renormalizable theory
is not necessarily guaranteed by the existence of a fixed point in the $\beta$
function and a second-order phase transition.
\vskip 0.2 true in
\proclaim 5. {\bf Concluding Remarks} \par
\vskip 0.2 true in
We have constructed an $SL(2,C)$ invariant theory of quantum gravity
by using a covariant spinor formalism and introducing an
$SL(2,C)$ Lagrangian density, which contains quadratic Yang-Mills-type
contributions. Problems with unitarity are avoided by invoking a non-linear
realization of the internal metric $s$ on the non-compact group $SL(2,C)$,
which is then realized linearly on $SU(2)$. By transforming to the gauge
$s=I$, the theory is spontaneously broken down to an $SU(2)$ invariant
theory and quantized on a lattice with the lattice spacing equal to the
Planck length. Although Lorentz invariance is broken in the gauge $s=I$, this
happens only in this particular gauge and, in general, the theory is fully
Lorentz and diffeomorphism invariant. In the ``unitary'' gauge $s=I$,
the Hamiltonian is shown to be bounded from below. A general gauge independent
analysis will maintain the positivity of the Hamiltonian as a general feature
of the theory. Einstein's locally Lorentz invariant and diffeomorphism
invariant theory will dominate at low energies and produce the standard
agreement with observational tests, because the extra degrees of freedom
related to the breaking of local Lorentz invariance in the gauge $s=I$ will
be undetectable at low energies. Thus, in this scheme, gravity in the gauge
$s=I$, becomes an $SU(2)$ invariant gauge theory, which is dominated by
Einstein's $SL(2,C)$ invariant gravity theory at large distances.

We then quantize the $SU(2)$ (or $SO(3)$) invariant theory
on a lattice by constructing the discretized action, which then produces
a functional generator $Z$ in terms of the discretized path integral.

Further work must be done to investigate the possible existence of a continuum
limit or, alternatively, the existence of a finite renormalizable version of
the
$SU(2)$ gauge invariant gravity theory using the finite lattice spacing
$a=G^{1/2}$. This program
is facilitated by the fact that an $SU(2)$ lattice calculation is not difficult
to perform using, for example, the heat kernel technique and
Monte Carlo simulation methods$^{24}$.

In contrast to the work of Ashtekar$^{3}$, in which a complex connection
is introduced based on a complex $SO(3,1;C)$ group structure, or an
$SL(2,C)$ connection which is self-dual, the present
theory leads to a real Hamiltonian, and there is no problem in defining
physical state vectors and inner products for a Hilbert space.

If we assume that a first-order phase transition occurs at high temperatures,
$T\sim T_c$, due to a non-vanishing vacuum expectation value:
$$
<\phi^{\dag}s\phi>_0=v^2,
\sectionit
$$
corresponding to a minimuum in the potential:
$$
V^\prime(\phi^{\dag}s\phi)=0,
\sectionit
$$
then the physical true vacuum will be spontaneously broken, due to the
breaking of local and diffeomorphism invariance. This will
correspond to a standard Higgs breaking of the true vacuum, which leads
to interesting consequences for the problem of time in quantum gravity$^{8}$,
early Universe cosmology$^{9}$ and
black hole evaporation and information loss phenomena$^{10}$. The matter
part of the Higgs mechanism associated with the matter fields $\phi$
will possess a physical particle spectrum free of ghost poles and tachyons,
because of the non-linear realization of the $\phi$ matter section of
the Lagrangian density (3.4) through the Hermitian internal metric $s$.

Finally, we should note that the present theory could have been formulated
in terms of real vierbeins $e^a_\mu$, defined by the spacetime metric:
$$
g_{\mu\nu}=e^a_\mu e^b_\nu \eta_{ab}.
\sectionit
$$
A Lagrangian density could be constructed which is invariant under $SO(3,1)$
or $SL(4,R)$ gauge transformations. Then unitarity for the scheme will be
guaranteed by non-linearly realizing these non-compact groups on a real
positive internal metric, $s_{ab}$. In the particular gauge $s=I$, local
Lorentz invariance will be broken down to $SO(3)$ for the $SO(3,1)$
theory and $O(4)$ for the $SL(4,R)$ theory. As in the $SL(2,C)$ gauge
theory, the Lagrangian density will be invariant under the full non-compact
gauge group through the extended internal metric theory.
\vskip 0.2 true in
{\bf Acknowledgements}
\vskip 0.2 true in
I thank M. Clayton, N. Cornish, L. Demopolous, and J. Greensite for
stimulating and helpful discussions. This work was supported by the Natural
Sciences and Engineering Research Council of Canada.
\vskip 0.2 true in
\centerline{\bf References}
\item{1.}{E. Alvarez, Rev. Mod. Phys. {\bf 61}, 561 (1989); C. Isham,
``Prima Facie Questions in Quantum Gravity'', Imperial
College preprint, Imperial/TP/93-94/1, 1993.}
\item{2.}{K. S. Stelle, Phys. Rev. D {\bf 16}, 953 (1977); Gen. Rel. and
Grav. {\bf 9}, 353 (1978); E. Sezgin and P. van Nieuwenhuizen, Phys. Rev.
D {\bf 21}, 3269 (1980); L. Smolin, Nucl. Phys. B {\bf 247}, 511 (1984);
C. Rovelli and L. Smolin, Nucl. Phys. B {\bf 331}, 80 (1990); Proceedings of
the Osgood Hill Conference on Conceptual Problems in Quantum Gravity,
eds. A. Ashtekar and J. Stachel, Birkh\"auser, Boston, 1991.}
\item{3.}{A. Ashtekar, Phys. Rev. D {\bf 36}, 1587 (1987);``Lectures on
Non-Perturbative Canonical Gravity'',
World Scientific Press, Singapore (1991); P. Peld\'an, ``Actions for Gravity,
with Generalizations: A Review'', Chalmers University preprint, G\"oteborg
ITP 93-13, 1993.}
\item{4.}{R. Capovilla, J. Dell, and T. Jacobson, Phys. Rev. Lett. {\bf 63},
2325 (1991); R. Capovilla, J. Dell, and L. Mason, Class. Quant. Grav.
{\bf 8}, 41 (1991).}
\item{5.}{See, for example, P. G. Lepage, ``From Action to Answers'',
Proceedings of the 1989 Advanced Institute in Elementary Particle Physics'',
eds., T. DeGrand and T. Toussaint, World Scientific, Singapore, p. 483 (1990).}
\item{6.}{B. Br\"ugmann, ``Loop Representations'', Max Planck Institute of
Physics preprint, MPI-Ph/93-94 (1993).}
\item{7.}{C. Rovelli, Class. Quan. Grav. {\bf 8}, 1613 (1991); Nucl. Phys.
B {\bf 405}, 797 (1993).}
\item{8.}{J. W. Moffat, Found. of Phys. {\bf 23}, 411 (1993).}
\item{9.}{J. W. Moffat, Int. J. Mod. Phys. D {\bf 2}, 351 (1993).}
\item{10.}{J. W. Moffat, ``Predictability in Quantum Gravity and Black Hole
Evaporation, University of Toronto preprint, December 1993, UTPT-93-31.}
\item{11.}{W. L. Bade and H. Jehle, Rev. Mod. Phys. {\bf 25}, 714 (1953). }
\item{12.}{R. Penrose, Ann. Phys. (N.Y.) {\bf 10}, 171 (1960); E. T. Newman
and R. Penrose, J. Math. Phys. {\bf 3}, 566 (1962).}
\item{13.}{M. Carmeli, E. Leibowitz, and N. Nissani, ``Gravitation: $SL(2,C)$
Gauge Theory and Conservation Laws'', World Scientific, Singapore, 1990.}
\item{14.}{T. T. Wu and C. N. Yang, Phys. Rev. D {\bf 13}, 3233 (1976).}
\item{15.}{J. P. Hsu and M. D. Xin, Phys. Rev. D {\bf 24}, 471 (1981).}
\item{16.}{The use of an internal metric in this kind of theory was first
proposed by K. Cahill, Phys. Rev. D {\bf 18}, 2930 (1978); {\bf 20}, 2636
(1979);
J. Math. Phys. {\bf 21}, 2676 (1980); Phys. Rev. D {\bf 26}, 1916 (1982).}
\item{17.}{B. Julia and F. Luciani, Phys. Lett. B {\bf 90}, 270 (1980).}
\item{18.}{J. E. Kim and A. Zee, Phys. Rev. D {\bf 21}, 1939 (1980).}
\item{19.}{J. Dell, J. L. de Lyra, and L. Smolin, Phys. Rev. D {\bf 34},
3012 (1986).}
\item{20.}{D. S. Popovi\'c, Phys. Rev. D {\bf 34}, 1764 (1986).}
\item{21.}{P. A. M. Dirac, Proc. R. Soc. London A {\bf 246}, 333 (1958);
``Lectures on Quantum Mechanics'', Belfer Graduate School of Science, Monograph
Series No. 2. Yeshiva University: New York, 1964; A. Hanson, T. Regge, and
C. Teitelboim, ``Constrained Hamiltonian Systems'', Academia Nazionale
Dei Lincei, Rome, 1976.}
\item{22.}{L. Smolin, Phys. Rev. D {\bf 30}, 2159 (1984).}
\item{23.}{K. G. Wilson, Phys. Rev. D {\bf 14}, 2455 (1974).}
\item{24.}{M. Creutz, ``Quarks, Gluons and Lattices'', Cambridge University
Press, Cambridge, 1983.}

\end